\documentclass[secnumarabic,amssymb, nobibnotes, aps, prd, 11pt, abstract]{revtex4-2}
\usepackage{tocloft}
\usepackage[dvips]{graphicx}
\usepackage{amsmath}
\usepackage{subcaption}
\usepackage{hyperref}
\usepackage{filecontents}
\hypersetup{colorlinks,linkcolor={red},citecolor={blue},urlcolor={red}}
\usepackage{setspace}
\usepackage{xcolor}
\usepackage{braket}
\pagecolor{white}
\usepackage{float}
\graphicspath{{./plots/}}

\usepackage{multirow}
\color{black}

\begin{document}

\singlespacing

	\title{Dynamical Timelike Entanglement Entropy in an Evaporating Schwarzschild--AdS Black Hole}

	\author{Digen Das$^{1,2}$}
	
	\email{$rs_digendas@dibru.ac.in$}

	\author{Prabwal Phukon$^{1,3}$}
	\email{$prabwal@dibru.ac.in$}
	
	\affiliation{$^1$Department of Physics, Dibrugarh University, Dibrugarh, Assam,786004.\\$^2$ Department of Physics, Sadiya College, Chapakhowa, Assam,786157.\\$^3$Theoretical Physics Division, Centre for Atmospheric Studies, Dibrugarh University, Dibrugarh,Assam,786004.\\}

	\begin{abstract}
Timelike entanglement entropy (tEE) has recently emerged as a novel probe of temporal quantum correlations in gravitational systems. Existing studies are largely restricted to static backgrounds. In this work we extend the construction of tEE to an evaporating Schwarzschild--AdS black hole. The evaporation is modeled through an effective Stefan--Boltzmann description of Hawking radiation coupled to an external absorptive bath, yielding a time-dependent horizon radius ($r_h(t)$) and surface gravity $\kappa(t)$. We derive, from the near-horizon Rindler structure of the evolving horizon, an adiabatic generalization of the static Kruskal construction, obtaining the accumulated thermal phase $\Phi(t)=\int_0^t\kappa(t')\,dt'$ as the natural replacement for the static phase $\kappa t$. The validity of this construction is governed by an explicit adiabatic parameter $\mathcal A(t)$, which we verify numerically remains small ($\lesssim0.012$) throughout the regime of interest. It vanishes exactly where the horizon crosses the critical radius $r_h=l/\sqrt3$ identified independently from the static thermodynamics. Using $\Phi(t)$, we construct a dynamical timelike entanglement entropy that continuously tracks the evaporation process and derive the corresponding dynamical Page-like times. Unlike the uniformly spaced Page-like times of the static geometry, evaporation induces non-uniform temporal spacing, together with a progressive phase delay and amplitude modulation of the oscillatory tEE. Because the dynamical entropy depends on the full accumulated history of $\kappa(t)$ rather than its instantaneous value alone, it retains a memory of the entire evaporation process. These results establish a first-principles dynamical framework for investigating temporal quantum correlations in evaporating black holes.
\end{abstract}
\maketitle	
\section{Introduction}\label{sec1}
The study of black holes from the perspective of thermodynamics has emerged as one of the most actively pursued areas in the field of theoretical physics. It bridges quantum mechanics, general relativity, and thermodynamics in a remarkable way. Black holes serve as the natural laboratory for the development as well as testing of different kinds of gravitational theory such as string theory \cite{rovelli2008loop} , loop quantum gravity \cite{grana2006flux} etc. One of the most fascinating properties of black hole, discovered by Hawking \cite{hawking1975particle, wald1975particle} is that it can evaporate through radiation commonly known as Hawking radiation. According to this an isolated black hole can lose mass and eventually it evaporates completely. However, this ground breaking result built a fundamental conflict in the field of theoretical physics. A black hole formed from pure matter state may evaporate completely through Hawking radiation which is thermal radiation and is described by a mixed state \cite{hawking1976breakdown}. So it implies that the black evaporation is from a pure state to a mixed state which contradicts the unitary principle of quantum mechanics. This violation of unitary principle is manifested in the monotonically increasing behaviour of Von Neumann Entropy of the emitted radiation, as suggested by Hawking in his semiclassical description of the black hole evaporation. This fundamental issue is commonly known as information paradox. A clear understanding of the information paradox may provide useful insights towards development of a quantum theory of gravity.
\par To make a consistent theory of black hole evaporation, Don Page suggested that the entanglement entropy of Hawking radiation should follow a characteristic behavior, now known as the Page curve \cite{page1993information, page2013time}. According to this, the entanglement entropy of Hawking radiation increases in early times, reaches a maximum value and then declines to zero at later times of evaporation. The time at which it reaches the maximum value is called as the Page time. Furthermore, the discovery of the AdS/CFT correspondence \cite{maldacena1999large} provides strong evidence for information conservation in black hole evaporation. According to this correspondence, there exist a duality between quantum gravity in Anti-de Sitter (AdS) spacetime and a conformal field theory defined on its boundary. 
\par Different theoretical approaches have been developed to understand how the Page curve appears in a theory that consistently combines quantum mechanics and gravity such as the firewall proposal \cite{almheiri2013black} and island formula \cite{penington2020entanglement, almheiri2019entropy}. The former argues that if information is preserved and hence the evaporation process is unitary in nature then the event horizon cannot be smooth. The smoothness of the event horizon follows from the equivalence principle of general relativity according to which, any freely falling observer experiences no dramatic physical effects while crossing the event horizon. 
\par On the other hand, the latter is generalization of an earlier framework \cite{ryu2006holographic} developed nearly two decades ago. In that work, the authors showed that the entanglement entropy S(R) of a region R and its complement in a conformal field theory (CFT) can be calculated from the dual Anti-de Sitter (AdS) geometry by determining the area of a minimal, or more generally extremal, surface in the bulk whose boundary coincides with that of R. By incorporating quantum corrections, the Ryu–Takayanagi (RT) prescription was further extended to the framework of quantum extremal surfaces (QES) \cite{barrella2013holographic, hubeny2007covariant, lewkowycz2013generalized, engelhardt2015quantum, faulkner2013quantum}. This generalized formulation provides a systematic approach for computing the entanglement entropy of Hawking radiation. The resulting prescription is commonly referred to as the island rule. According to the  island rule the entanglement entropy of Hawking radiation is determined by the following expression:
\begin{equation}
\begin{split}
     S_\text{Rad} &= \text{min}(\text{ext}[S_\text{gen}]) \\
    S_\text{gen} &= \frac{Area (\partial I)}{4 G} + S_\text{field} (R \cup I) \label{eq1}
\end{split}
\end{equation}
where $S_{gen} $\ is the generalized entropy, \textit{R} denotes the radiation region and \textit{I} represents the island region within the black hole interior, $\partial I$ specifies the boundary of the island, whereas $S_{field}$ denotes the entanglement entropy of quantum fields in the combined region $R \cup I$. The location of the island can be found by extremising the above equation with respect to position and time. It has been well established that using the island paradigm, one can obtain the Page curve consistent with unitary of quantum mechanics. By implying the island rule, it has been found that the entanglement entropy of Hawking radiation increases at early stages prior to the Page time. At late times, island emerges and contributes to the entanglement entropy of the radiation and the entanglement entropy declines to a constant value at later times\cite{hashimoto2020islands, almheiri2020entanglement, anegawa2020notes, karananas2021islands, bak2021unitarity, tong2024island, yu2022island, lin2024entanglement}.
\par Despite these significant developments, the island formula still faces some limitations.  At present, the framework is rigorously established primarily within the context of the AdS/CFT correspondence and its applicability to black holes in asymptotically flat or de-Sitter spacetime remains unclear\cite{hao2025non, goswami2022small}. Besides, it has been reported that no island emergences in Liouville black holes \cite{li2021island} and hence it leaves scope for searching complimentary approaches that may provide interesting insights regarding the same.
\par Another interesting concept in the field of quantum entanglement is timelike entanglement entropy that has recently attracted considerable attention\cite{olson2012extraction}. The basic difference between the conventional entanglement entropy and timelike entanglement entropy is that timelike entanglement describes correlations between states of the same physical system evaluated at different points in time \cite{nowakowski2018entangled}. This idea has been extended using the AdS/CFT correspondence that  leads to the notion of holographic timelike entanglement entropy \cite{harper2023timelike}. As the regions considered in this case are separated in a temporal way, the corresponding density matrix associated with the same is non hermitian in nature that makes the timelike entanglement entropy a complex valued quantity. By performing an appropriate wick rotation we can calculate the timelike entanglement entropy and it is commonly interpreted as pseudo entropy \cite{nakata2021new, doi2023pseudoentropy, das2024bulk}. The imaginary component of the pseudo entropy may describe emergent time, generalizing the concept of emergent spatial geometry derived from quantum entanglement \cite{harper2023timelike}. Using the Rindler method, the timelike entanglement entropy can be interpreted as the thermal entropy of the conformal field theory (CFT) after a Rindler transformation, up to an additive constant\cite{he2024holographic}. Very recently, the timelike entanglement entropy of Hawking radiation has been studied across a range of black hole solutions with fixed surface gravity, revealing periodic recurrences governed by $\kappa$ \cite{ladghami2026timelikeentanglemententropyhawking}. The present work differs in letting the surface gravity itself evolve under Hawking evaporation, and derives the resulting dynamical timelike entanglement entropy from first principles.
\par
The primary objective of this work is to investigate the evolution of temporal quantum correlations during black-hole evaporation through the framework of timelike entanglement entropy. Rather than introducing the required generalization of the static construction by analogy, we derive it from first principles, exploiting the near-horizon Rindler structure of the evolving horizon and its adiabatic generalization to a time dependent surface gravity; the resulting construction is controlled by an explicit adiabatic parameter, which we verify remains small throughout the regime of interest. Using this construction, we show that the resulting dynamical timelike entanglement entropy is intrinsically history dependent: unlike the static case, where the entropy is fixed by the instantaneous surface gravity alone, the dynamical entropy at any instant encodes the full accumulated thermal history of the evaporating black hole. This history dependence manifests concretely as a non-uniform spacing of the Page-like times and a progressive modulation of the oscillation amplitude, in contrast to the uniformly spaced, fixed amplitude oscillations of the static geometry. Rather than replacing the conventional Page curve or the island prescription, our approach complements these frameworks, offering a probe of an evaporating black hole's temporal correlation structure that is sensitive not only to its instantaneous thermodynamic state but to its entire prior evolution, and provides a framework for investigating this history dependence in more general dynamical gravitational backgrounds.

The remainder of this paper is organized as follows. In Sec. II, we briefly review the holographic aspects of timelike entanglement entropy and Hawking Radiation. In Sec. III we introduce an effective Stefan--Boltzmann
description of Hawking evaporation, coupled to an external absorptive bath, and determine the resulting time dependence of the horizon radius and surface gravity. In Sec. IV, we derive the accumulated thermal phase governing the evaporating background from first principles, by exploiting the near-horizon Rindler structure of the evolving horizon and its adiabatic generalization to a time-dependent surface gravity.
In Sec. V, we combine the evaporation dynamics of Sec. III with the accumulated thermal phase of Sec.~IV to construct the dynamical timelike entanglement entropy of the evaporating
Schwarzschild--AdS black hole, and derive the associated dynamical Page-like times. Finally, in
Sec. VI, we summarize our results and discuss their limitations together with directions for
future work.

\section{Timelike Entanglement Entropy and Hawking Radiation}\label{sec2}
The entanglement entropy in \textit{d+1} dimensional conformal field theory ($CFT_{d+1}$) can be computed via the AdS/CFT correspondence \cite{ryu2006aspects}. In the framework of the AdS/CFT correspondence, the holographic entanglement entropy corresponding to a boundary subregion A is determined by the area of an extremal surface 
in the bulk spacetime whose boundary coincides with that of the subregion A, given by - 
\begin{equation}
    S_A = \frac{Area (\gamma_A)}{4 G}
\end{equation}
where G is Newton's constant. This relation is commonly referred to as the Ryu–Takayanagi prescription. A remarkable insight has emerged from this idea, that the emergent spatial geometry of Anti-de Sitter (AdS) spacetime is deeply connected to quantum entanglement in the dual conformal field theory. This connection suggests that spacetime geometry itself may arise from underlying quantum informational structures. Motivated by this idea, the authors of \cite{harper2023timelike} have raised the intriguing possibility that not only spatial coordinates, but also the time coordinate, may emerge from quantum information theoretic properties. In this context, a new quantity known as the timelike entanglement entropy (tEE) has been proposed. The timelike entanglement entropy can be calculated through an analytic continuation of the standard entanglement entropy to subsystems defined over timelike regions or equivalently through an appropriate wick rotation\cite{harper2023timelike}. Furthermore, within the framework of  AdS/CFT correspondence, the imaginary part of the timelike entanglement entropy has been argued to be related to the area of a timelike extremal surface \cite{harper2023timelike}. As stated earlier, this complex valued timelike entanglement entropy is also known as pseudo entropy as introduced in \cite{nakata2021new}.
\par Consider a system with total Hilbert space $\mathcal{H}_{total}$ and divide the system into two subsystems A and B. Then  decomposing $\mathcal{H}_{total}$ into those of A and B i.e. $\mathcal{H}_{total} = \mathcal{H}_{A} \otimes \mathcal{H}_{B}$, the pseudo entropy can be defined as follows - 
\begin{equation}
    S_A = -\text{Tr}[\rho_A \log \rho_A]
\end{equation}
where, $\rho_A$ is the reduced density matrix and can be written as - 
\begin{equation}
    \rho_A =\mathrm{Tr_B} \Bigg[\frac{|\psi\rangle \langle \phi|}{\langle \phi | \psi \rangle}\Bigg]
\end{equation}
 with the states $\ket{\psi}$ and $\ket{\phi}$ being two different pure states defined on $\mathcal{H}_{total}$.
 \par Now considering a two dimensional conformal field theory, we proceed to study timelike entanglement entropy. We define our subsystem A with an interval $I_{MN}$ where the end points M and N has the co-ordinates $(t_M, x_M)$ and $(t_N, x_N)$ respectively. Now, in terms of geodesic distance between P and Q, we can define the Von Neumann entropy of this sub-system with the following formula-
 \begin{equation}
     S_A = \frac{c}{3} \log \Bigg(\frac{\sqrt{(x_M-x_N)^2-(t_M-t_N)^2}}{\epsilon}\Bigg)
 \end{equation}
 where c is the central charge and $\epsilon$ is the ultraviolet cutoff. For temporally separated regions, $x_M = x_N$, then the timelike entanglement in terms of geodesic distance is given by - 
\begin{equation}
    S_A = \frac{c}{3} \log \Bigg(\frac{t_M-t_N}{\epsilon}\Bigg) + i\frac{\pi c}{6}
\end{equation}
\par 
Hawking radiation, as introduced in \cite{hawking1975particle} arises as a quantum effect in curved spacetime. In that context, black holes emit thermal radiation with a temperature determined by the surface gravity at the event horizon. This phenomenon implies that black holes behave as a thermodynamics system with temperature and entropy. In semiclassical theory of gravity, the emitted Hawking quanta are entangled with interior degrees of freedom of the black hole which leads to a gradual increase in the entanglement entropy of the radiation during the evaporation process. This behavior gives rise to the black hole information paradox as the thermal nature of Hawking radiation appears to evolve an initially pure quantum state into a mixed state that violates unitarity. Within the conventional framework, the entanglement entropy of Hawking radiation is usually analyzed through spatial partitions of the quantum system. This is encoded in the Page curve via island prescription \cite{hashimoto2020islands,lin2024entanglement,wang2024entanglement, almheiri2021entropy}. However, recent developments in holography and pseudo entropy suggest that quantum correlations may also be fundamentally temporal in nature \cite{harper2023timelike}. As stated earlier, Timelike entanglement entropy characterizes quantum correlations between states separated along the time direction rather than across spatial subregions. This motivates the possibility that temporal correlations associated with  radiation may provide complimentary information regrading the black hole evaporation. From this perspective, timelike entanglement entropy provides a natural framework for probing the dynamical and thermal structure of Hawking radiation.  It may reveal new aspects of information flow and non equilibrium quantum gravitational dynamics in black hole spacetimes.

\section{Evaporating Schwarzschild AdS Black Hole} \label{sec3.1}

To illustrate the proposed framework, we begin with the Schwarzschild black hole in Anti de-Sitter space. This system serves as the simplest and most analytically tractable solution of Einstein’s field equations. It provides an ideal set up  for exploring the fundamental aspects of black hole thermodynamics and quantum information theoretic quantities. The spacetime geometry of the Schwarzschild AdS black hole is described by the metric - 
\begin{equation}
ds^2=-f(r)\,dt^2+\frac{dr^2}{f(r)}+r^2 d\theta^2 + r^2 \sin^2\theta  d\phi^2
\end{equation}
with the metric function is given by
\begin{equation}
f(r)=1-\frac{2MG}{r}+\frac{r^2}{l^2}
\end{equation}
where, event horizon $r_h$ can be calculated by solving the equation $f(r_h) = 0$, M is the mass of the black hole, G is Newton's constant and l is the AdS length. For simplicity we will use natural unit in the rest of this paper and set \textit{l}=1 for numerical calculations.
\par Similar to asymptotically flat black holes, Schwarzschild-AdS black holes emit Hawking radiation as a consequence of quantum field theoretic effects in curved spacetime. The Hawking temperature is determined by the surface gravity at the event horizon and is given by -
\begin{equation}
T_H = \frac{\kappa}{2\pi} = \frac{f'(r_h)}{4\pi}= \frac{1}{4\pi} \left(\frac{1}{r_h} +\frac{3r_h}{l^2} \right) , \quad \kappa = \frac{1}{2r_h} \bigg(1+ \frac{3r_h^2} {l^2}\bigg)
\end{equation}
The exact computation of entanglement entropy in higher-dimensional black hole spacetime is generally a highly nontrivial problem. However, in the asymptotic region far from the black hole, the dominant contribution to Hawking radiation arises from the low angular momentum sector, particularly the s-wave mode. Consequently, one may effectively neglect the angular dependence of the spacetime and reduce the problem to an effective two-dimensional geometry described by the $(t,r)$ sector of the metric \cite{hashimoto2020islands,guo2023page}.Within this effective two-dimensional description, the temporal correlation structure associated with Hawking radiation can be analyzed in a tractable manner through Euclidean continuation and timelike geodesic separations. We consider a Hawking radiation region defined by the interval $I_{MN}$, where the endpoints are $M=(t_M,x_M)$ and $N=(t_N,x_N)$. For an effectively two-dimensional geometry obtained under the s-wave approximation, the entanglement entropy of a radiation interval can be related to the geodesic separation between the corresponding boundary points. Consequently, the von Neumann entropy associated with the Hawking radiation region may be identified with the entropy of a two-dimensional conformal field theory defined on the same interval. For a two-dimensional conformal field theory, the entanglement entropy takes the form \cite{calabrese2004entanglement, lin2024entanglement}- 
\begin{equation}
S_R=S_{\mathrm{CFT}}=\frac{c}{3}\log\left(\frac{d(M,N)}{\epsilon}\right) 
\end{equation}
where c is the central charge, $d (M,N)$ is the geodesic distance between the end points of the interval $I_{MN}$ and $\epsilon$ is the UV cutoff.
\par As stated earlier, for a temporally separated region, $x_M = x_N$. In order to analyze such timelike correlations, we analytically continue the Schwarzschild-AdS geometry to Euclidean signature by performing a  Wick rotation  $t\rightarrow i\tau $ which transforms the Schwarzschild metric in the form \cite{battista2022geodesic} - 
\begin{equation}
ds_E^2=f(r)\, d\tau^2+\frac{dr^2}{f(r)}+r^2 d\Omega_2^2 
\end{equation}
where, \begin{equation}
d\Omega_2^2=d\theta^2+ \sin^2\theta\, d\phi^2 .
\end{equation}
To investigate the temporal correlation structure associated with Hawking radiation, it is convenient to introduce coordinates that remain regular at the event horizon. For this purpose, we first define the tortoise coordinate $(r_*)$ through \cite{lin2024entanglement}

\begin{equation}
r_*=\int \frac{dr}{f(r)}
\end{equation}

where \textit{f(r)} is the Schwarzschild-AdS metric function. Near the event horizon $(r=r_h)$, the metric function can be expanded as

\begin{equation}
f(r)\simeq f'(r_h)(r-r_h)
=2\kappa (r-r_h)
\end{equation}

where $(\kappa)$ denotes the surface gravity of the black hole. Consequently, the tortoise coordinate exhibits the asymptotic behavior

\begin{equation}
r_*\simeq \frac{1}{2\kappa}
\ln |r-r_h|
\end{equation}

which diverges logarithmically at the horizon. This divergence can be removed by introducing suitable Kruskal coordinates. We define the Euclidean Kruskal coordinates as 

\begin{equation}
U=-e^{-\kappa(i\tau-r_*)},
\qquad
V=e^{\kappa(i\tau+r_*)}
\end{equation}

These coordinates provide a regular description of the near-horizon geometry and naturally encode the thermal periodicity associated with Hawking radiation. Within the effective two-dimensional description and the s-wave approximation, the geodesic distance between the end points M and N can be written as

\begin{equation}
d^{2}(M,N)= \Omega(M)\Omega(N)
\left(U_M-U_N\right)
\left(V_N-V_M\right)
\end{equation}

where the conformal factor is given by \cite{lin2024entanglement} - 
\begin{equation}
  \Omega(r) = \frac{1}{\kappa e^{\kappa r_*(r)}}  
  \end{equation}
Now using the above equations, it is straightforward to calculate the timelike entanglement entropy of effective quantum fields in Schwarzschild AdS space time as - 
\begin{equation}
    S_R (t)= \frac{c}{3} \log\bigg[\frac{4 \sin^2(\frac{\kappa t}{2})}{\kappa^2 \epsilon^2} \bigg] \label{tEE:Static}
\end{equation}
where $t = \tau_M - \tau_N$.  Unlike the conventional entanglement entropy obtained from spatially separated regions, the above quantity characterizes the quantum correlations between radiation states separated in a timelike manner. One of the notable results is that  the entropy depends explicitly on the surface gravity $(\kappa)$, indicating that the temporal correlation structure is governed by the thermal properties of the black hole. Secondly, the appearance of the trigonometric factor originates from the Euclidean periodicity of the Schwarzschild-AdS geometry. This reflects the thermal nature of Hawking radiation. As a consequence, the timelike entanglement entropy exhibits a periodic oscillatory behavior with a characteristic recurrence timescale which is directly determined by the inverse surface gravity. The periodicity of the entropy may be obtained directly from the argument of the trigonometric function, yielding

\begin{equation}
T_{\rm rec}= \frac{2\pi}{\kappa}.  
\end{equation}

This recurrence timescale coincides with the Euclidean thermal period of the black hole. Hence it provides a direct connection between the temporal entanglement structure and the underlying thermodynamic properties of the horizon. From the above equation, it is clear that the first maximum occurs at $t_P=\pi /\kappa$. In the static Schwarzschild-AdS background, this time scale represents the first  recurrence of the timelike entanglement entropy. Motivated by its role as the first entropy maximum, we refer to this quantity as a Page-like time scale. In the evaporation scenario discussed later, this quantity acquires a dynamical interpretation. It should also be noted that, the subsequent maxima occur periodically with separation $t_P= \frac{\pi}{\kappa}+ \frac{2n\pi}{\kappa} $, where \textit{n} is an integer that labels the subsequent Page-like time. This indicates that the temporal correlation structure exhibits a hierarchy of recurring Page-like times that are governed by the Hawking temperature of the black hole. 

Unlike asymptotically flat black holes, Hawking radiation emitted by a Schwarzschild-AdS black hole does not escape to infinity. Instead, the timelike AdS boundary acts like an effective reflecting surface. This  causes the emitted radiation to return to the black hole and establishing thermal equilibrium. As a result, an eternal Schwarzschild-AdS black hole does not undergo sustained evaporation. To investigate the evolution of the timelike entanglement entropy in this system, we couple the black hole to an external non-gravitational bath. This absorbs the outgoing Hawking radiation. The black hole is then allowed to lose energy through Hawking emission. In this way mass, horizon radius, and surface gravity becomes time dependent.

To model this process, we adopt an effective Stefan-Boltzmann description of Hawking radiation and approximate the mass loss rate by

\begin{equation}
\frac{dM}{dt}=-\alpha A_h T_H^4, \label{SBeq}
\end{equation}

where$ (A_h=4\pi r_h^2)$ is the horizon area and $(T_H)$ is the Hawking temperature. For simplicity we neglect grey body corrections and take $\alpha=\pi^2/60$ in natural units \cite{giddings2016hawking, chaudhary2024effects}. Now, the mass of the black hole can be written as 

\begin{equation}
M(r_h)= \frac{r_h}{2l^2} \left(l^2+r_h^2\right)
\end{equation}

Having considered the evaporation model, we now determine the time evolution of the black hole horizon. Since the black hole mass is expressed as a function of the horizon radius, we can write 
\begin{equation}
\frac{dM}{dt} =\frac{dM}{dr_h} \frac{dr_h}{dt}.
\end{equation}

and 
\begin{equation}
\frac{dM}{dr_h} =\frac{1}{2}\left(1+\frac{3r_h^2}{l^2} \right).
\end{equation}
Substituting this expression in Equation \ref{SBeq},
we obtain the evolution equation governing the horizon radius,
\begin{equation}
\frac{dr_h}{dt}
=-\frac{\alpha\left(l^2+3r_h^2\right)^3}{32\pi^3l^6r_h^2}.
\label{eq:drdt}
\end{equation}

Equation (\ref{eq:drdt}) constitutes a differential equation describing the evaporation of the Schwarzschild--AdS black hole. We solve the equation numerically with the initial condition 
\begin{equation}
r_h(0)=2.
\end{equation}

Substituting the numerical solution for $r_h(t)$ into the expression for the surface gravity,
\begin{equation}
\kappa(t)
= \frac{1}{2r_h(t)}+\frac{3r_h(t)}{2l^2}, \label{eq:k(t)}
\end{equation}
yields the complete time dependence of the surface gravity. This quantity serves as the fundamental input for evaluating the accumulated thermal phase introduced in the following section and, consequently, the dynamical timelike entanglement entropy.
\par 
We first consider a naive, instantaneous estimate of the Page-like timescale, obtained by evaluating the static formula $t_P=\dfrac{\pi}{\kappa}$ pointwise at each horizon radius. This coincides with the properly accumulated Page-like time of Sec.~V only when $\kappa$ has been
approximately constant up to that instant. Evaluating this instantaneous estimate along the evaporation trajectory, the naive Page-like timescale becomes a time-dependent quantity,
\begin{equation}
t_P(r_h)= \frac{\pi}{\kappa(r_h)}=
\frac{2\pi r_h l^2}{l^2+3r_h^2}. \label{eq30}
\end{equation}
It is clear from the above equation that the behavior of this instantaneous estimate is non-monotonic. Both terms dominate in different horizon-size regimes. For small black holes ($r_h\ll l$), the Schwarzschild-like contribution, $\kappa\simeq\frac{1}{2r_h}$, dominates, and in this limit the black hole thermodynamics closely resembles that of an asymptotically flat Schwarzschild black hole. On the other hand, for large black holes ($r_h\gg l$), the AdS contribution becomes dominant, $\kappa\simeq\frac{3r_h}{2l^2}$. By extremising Eq.~\eqref{eq30} with respect to the horizon radius, one finds the critical horizon radius $r_h=l/\sqrt3$, at which this instantaneous estimate attains its maximum value, $t_P^{\rm max}=\pi l/\sqrt3$. This behavior highlights the interplay between the two contributions to the surface gravity and demonstrates that the temporal structure of timelike entanglement undergoes a qualitative transition as the evaporating black hole evolves from the large-black-hole phase to the small-black-hole phase. Section~V refines this instantaneous estimate into the exact accumulated phase condition, Eq.~\eqref{eq:dynamicPage}. This critical radius reappears in Sec.~IV, where it coincides with a local minimum of the
adiabatic parameter governing the validity of the near-horizon construction used there.

     \begin{figure}[t]
    \centering
    \begin{subfigure}[b]{0.45\textwidth}
        \centering
        \includegraphics[width=\textwidth]{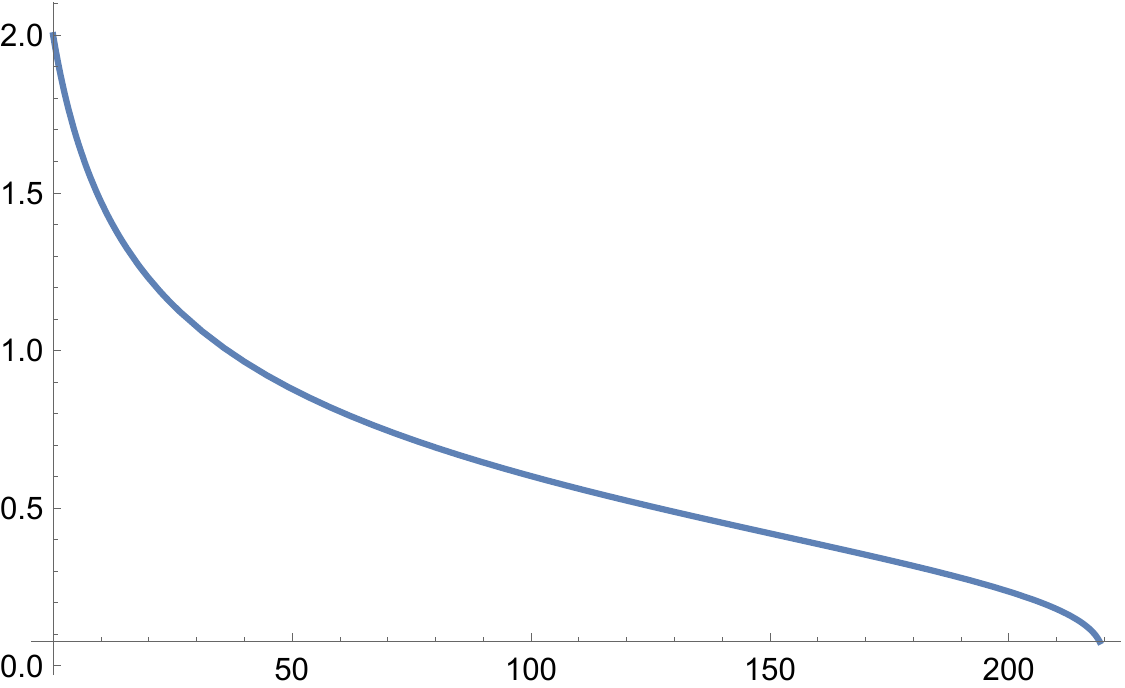}
        \caption{}
        \label{f1a}
    \end{subfigure}
    \hspace{0.03\textwidth}
    \begin{subfigure}[b]{0.45\textwidth}
        \centering
        \includegraphics[width=\textwidth]{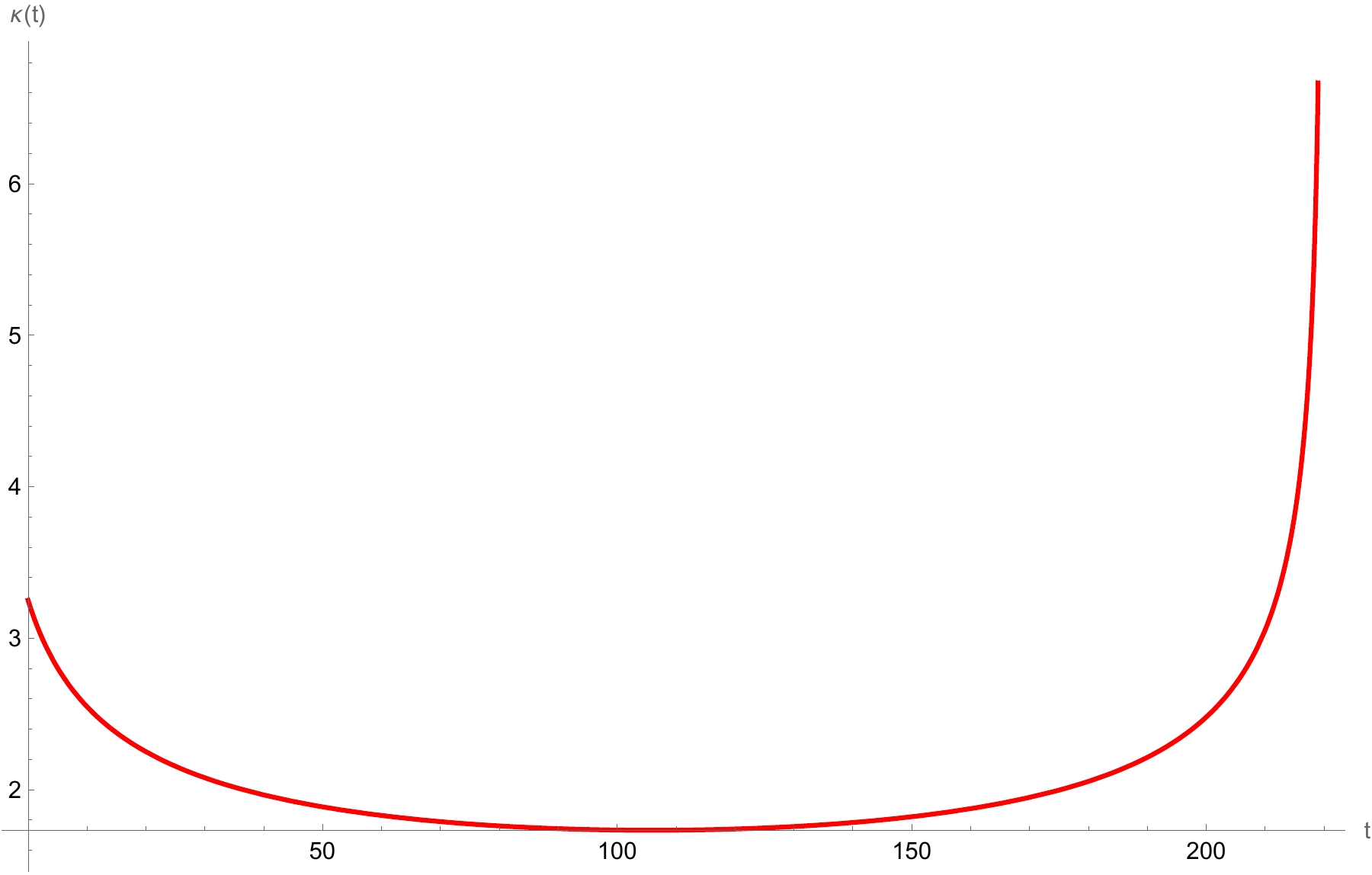}
        \caption{}
        \label{f1b}
    \end{subfigure}
    {\caption{(a) The variation of event horizon radius for a Schwarzschild AdS evaporating black hole (b) The variation of surface gravity $\kappa(t)$ for a Schwarzschild AdS evaporating black hole.}}
    \label{Plot1}
\end{figure}
\section{Near-horizon Rindler structure and its adiabatic generalization}\label{sec4}

The expression for the static timelike entanglement entropy discussed in  Sec.~II relies on the existence of
regular Kruskal-type coordinates covering the horizon. It is constructed from the constant surface
gravity $\kappa$ of the static Schwarzschild--AdS background. However, during evaporation the surface is no longer time independent and the construction of Sec.~II has to be generalized accordingly. For this, we derive it from the
near-horizon geometry directly, exploiting the fact that the vicinity of any non-degenerate Killing horizon is locally isometric to flat Rindler space. This provides a controlled route to the accumulated thermal phase employed throughout this work, together with an explicit statement of the approximation on which it rests.

In the static Schwarzschild--AdS geometry, expansion of the metric function about the horizon
shows that the near-horizon region is locally described by the Rindler metric,
\begin{equation}
ds^2=-\kappa^2\rho^2\,dt^2+d\rho^2 ,
\label{eq:rindler}
\end{equation}
where $\rho$ denotes proper distance and $\kappa$ is the (constant) surface gravity. A world line of fixed $\rho=\rho_0$ describes a uniformly accelerated observer with proper acceleration $1/\rho_0$; the horizon itself corresponds to the limit $\rho_0\to1/\kappa$. The associated Rindler--Minkowski transformation,
\begin{equation}
T=\rho\sinh(\kappa t) , \qquad X=\rho\cosh(\kappa t) ,
\end{equation}
yields the familiar null coordinates
\begin{equation}
U=T-X=-\rho\,e^{-\kappa t} , \qquad V=T+X=\rho\,e^{\kappa t} .
\label{eq:staticUV}
\end{equation}

\subsection{The Evaporating Case: An Adiabatic Ansatz}

\begin{figure}
    \centering
    \includegraphics[width=0.5\linewidth]{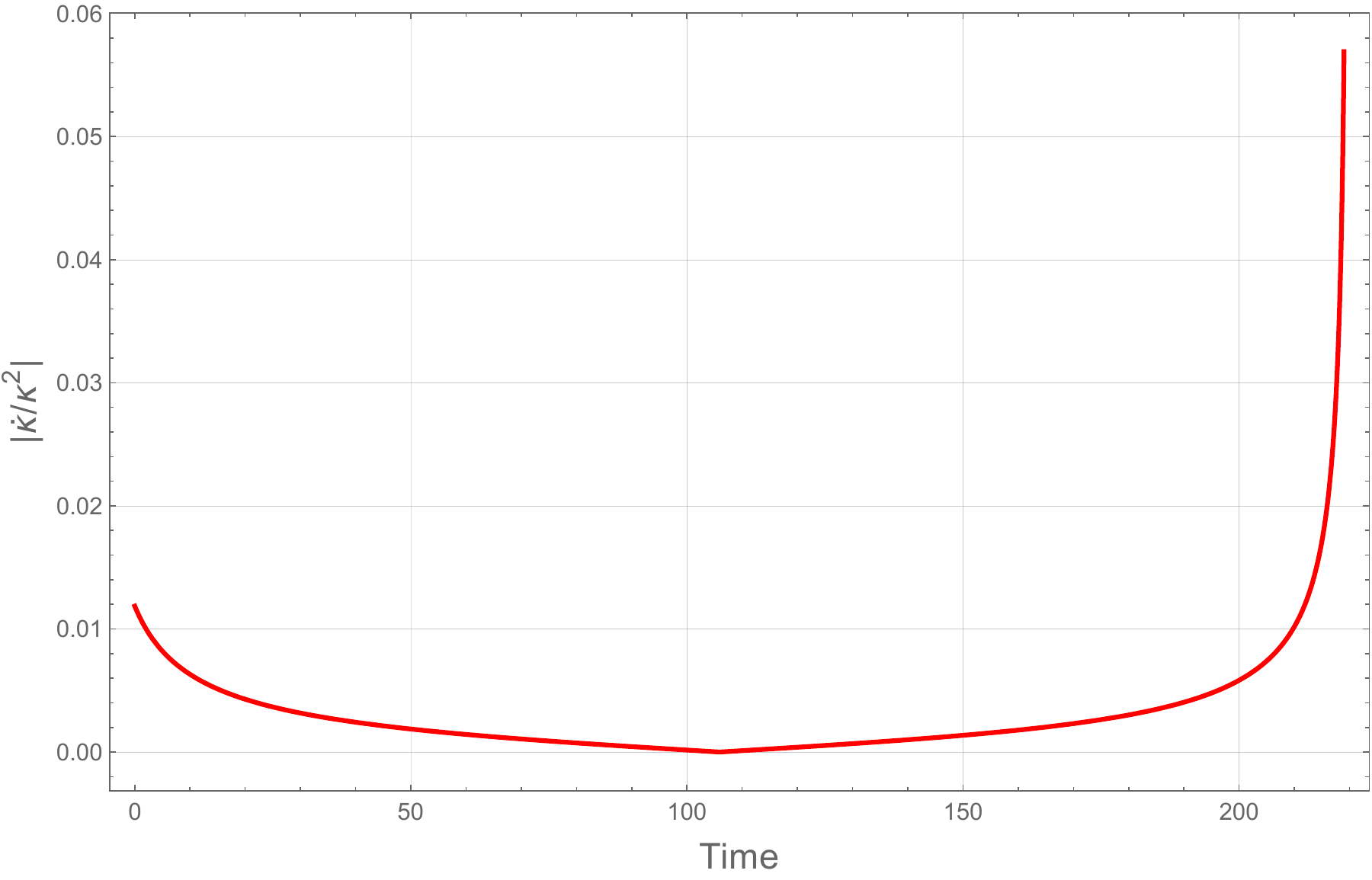}
    \caption{
The adiabatic parameter $\mathcal A(t)=|\dot\kappa(t)|/\kappa(t)^2$ as a function of time, computed from the numerical solution $r_h(t)$ of Eq.~(25). $\mathcal A(t)$ remains small ($\lesssim0.012$) over most of the evolution, attains a local minimum approaching zero at $t\approx106$ where the horizon crosses the critical radius $r_h=l/\sqrt3$ identified in Sec.~III, and grows to a maximum $\mathcal A_{\max}\approx0.057$ only in the terminal stage of evaporation as $r_h\to0$.}
    \label{Adiabetic plot}
\end{figure}

Under evaporation, the horizon radius as well as the surface gravity becomes time dependent, $\kappa\to\kappa(t)$. We model the near-horizon geometry at each instant $t$, by the
instantaneously Rindler form
\begin{equation}
ds^2 = -\kappa(t)^2\rho^2\,dt^2 + d\rho^2.
\label{eq:rindlerdyn}
\end{equation}
This expression is an ansatz, not a solution of the semiclassical Einstein equations: the true
near-horizon metric of a backreacted, radiating spacetime generically contains additional
$t$--$\rho$ cross terms sourced by the energy flux responsible for the mass loss, all of which are
omitted in Eq.~\eqref{eq:rindlerdyn}. The validity of this omission is controlled by the adiabatic
parameter defined by
\begin{equation}
\mathcal{A}(t) \equiv \frac{|\dot\kappa(t)|}{\kappa(t)^2} ,
\label{eq:adiabaticparameter}
\end{equation}
which measures the fractional change of the surface gravity over its own characteristic thermal time scale, $1/\kappa(t)$. The construction below is to be understood as valid on the domain where
$\mathcal{A}(t)\ll1$. Fig. \ref{Adiabetic plot} shows variation of $\mathcal A$ with time.

\subsection{Generalized Rindler--Minkowski Transformation}

Granting the ansatz \eqref{eq:rindlerdyn}, we define the accumulated rapidity
\begin{equation}
\theta(t) = \int_0^t \kappa(t')\,dt' ,
\label{eq:rapidity}
\end{equation}
and generalize the Rindler--Minkowski transformation by promoting the constant rapidity $\kappa t$
to $\theta(t)$,
\begin{equation}
T = \rho\sinh\theta(t) , \qquad X = \rho\cosh\theta(t) .
\label{eq:dynamicRindler}
\end{equation}
By differentiating the above equations, one finds $dT=\sinh\theta\,d\rho+\rho\dot\theta\cosh\theta\,dt$ and
$dX=\cosh\theta\,d\rho+\rho\dot\theta\sinh\theta\,dt$, from which we can find the induced line element by direct substitution:
\begin{equation}
-dT^2+dX^2
= d\rho^2 - \rho^2\dot\theta(t)^2\,dt^2
= d\rho^2 - \kappa(t)^2\rho^2\,dt^2 ,
\label{eq:exactreduction}
\end{equation}
where we have used $\dot\theta(t)=\kappa(t)$. Equation~\eqref{eq:exactreduction} reproduces the ansatz metric \eqref{eq:rindlerdyn} exactly. For arbitrary $\kappa(t)$ , the transformation \eqref{eq:dynamicRindler} is  an exact diagonalization of \eqref{eq:rindlerdyn}. 
The single physical approximation in the construction is the adoption of the ansatz \eqref{eq:rindlerdyn} itself, quantified by $\mathcal A(t)$. The corresponding null coordinates are obtained directly using the properties of hyperbolic function from Eq.~\eqref{eq:dynamicRindler}.
\begin{align}
U = T-X &= -\rho\left[\cosh\theta(t)-\sinh\theta(t)\right] = -\rho\, e^{-\theta(t)} ,
\label{eq:Uderiv}\\
V = T+X &= \rho\left[\cosh\theta(t)+\sinh\theta(t)\right] = \rho\, e^{\theta(t)} .
\label{eq:Vderiv}
\end{align}
By substituting $\theta(t)$ from Eq.~\eqref{eq:rapidity} into Eqs.~\eqref{eq:Uderiv}--\eqref{eq:Vderiv}, we get the
generalized null coordinates
\begin{equation}
U = -\rho\exp\!\left[-\int_0^t\kappa(t')\,dt'\right] ,
\qquad
V = \rho\exp\!\left[\int_0^t\kappa(t')\,dt'\right] ,
\label{eq:dynamicUV}
\end{equation}
It is also clear that the above equations reduce to the static result given by Eq.~\eqref{eq:staticUV} in the limit of constant $\kappa$.

\subsection{Thermality and the accumulated thermal phase}

The coordinate transformation derived above is purely geometric. This does not imply that the corresponding quantum state is thermal by itself. The physical interpretation follows from the response of an observer following the generalized Rindler trajectory \ref{eq:dynamicRindler}. For non-uniformly accelerated trajectories, Barbado and Visser~\cite{barbado2012unruh} showed that, in the adiabatic regime, the detector response is approximately Planckian with an instantaneous Unruh temperature $T(t)=\kappa(t)/2\pi$. 
Their adiabatic expansion, organized in powers of  the adiabatic parameter and its higher-order analogues shows that deviations from exact thermality are suppressed by the adiabatic parameter $\mathcal A(t)$. Thus, the thermality of the generalized Rindler trajectory relies on the same quasi-static assumption under which the near-horizon ansatz \eqref{eq:rindlerdyn} is expected to remain valid.
Motivated by this geometric construction and its approximate thermal interpretation, we identify
\begin{equation}
\Phi(t) \equiv \theta(t) = \int_0^t\kappa(t')\,dt' ,
\label{eq:accumulatedphase}
\end{equation}
as the dynamical generalization of the static phase $\kappa t$ and use it in the construction of the dynamical timelike
entanglement entropy.

\section{Dynamical timelike entanglement entropy} \label{sec5}

\begin{figure}
    \centering
    \includegraphics[width=0.5\linewidth]{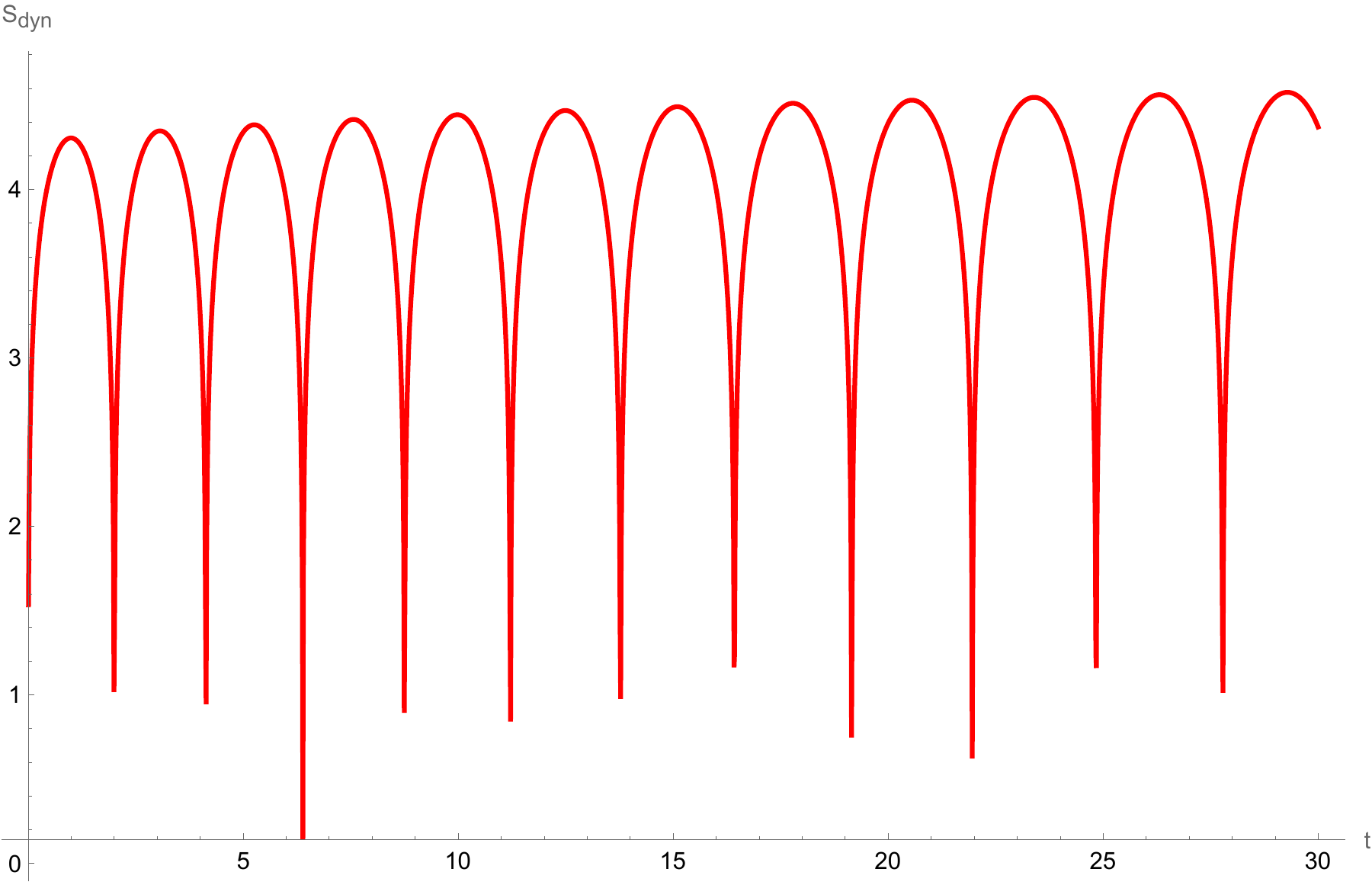}
    \caption{Dynamical timelike entanglement entropy $S_R^{\rm dyn}(t)$, as a function of time for the evaporating Schwarzschild--AdS black hole. The oscillatory behavior characteristic of the static case persists, but with maxima that become progressively delayed and enhanced in amplitude as the black hole evaporates.}
    \label{dynamic tEE}
\end{figure}

The time evolution of the horizon, obtained in Sec.~ III from the Stefan--Boltzmann mass-loss law
coupled to the external bath of Ref.~\cite{almheiri2020page}, fixes the surface gravity $\kappa(t)$ through Eq.~(\ref{eq:k(t)}). Section~IV showed that, under the near-horizon
adiabatic ansatz \eqref{eq:rindlerdyn}, valid on the domain $\mathcal A(t)\ll1$, the static phase
$\kappa t$ entering the Kruskal coordinates of Sec.~ II is replaced by the accumulated thermal
phase given by Eq (\ref{eq:accumulatedphase}). With these two ingredients in hand viz. $\kappa (t)$ and $\Phi(t)$
we can now write the dynamical
timelike entanglement entropy of the evaporating Schwarzschild--AdS black hole. Substituting
$\Phi(t)$ for the static phase $\kappa t$ in Eq.~(\ref{tEE:Static}),
\begin{equation}
S_R^{\rm dyn}(t) =
\frac{c}{3}\log\!\left[\frac{4\sin^2\!\left(\Phi(t)/2\right)}{\epsilon^2\kappa(t)^2}\right] ,
\label{eq7}
\end{equation}
The above equation (\ref{eq7}) provides a dynamical generalization of the timelike entanglement entropy. In case of static result, the oscillations occur with a constant frequency and fixed amplitude. On the other hand, the evolving surface gravity modifies both the phase and the overall magnitude of the entanglement entropy. The accumulated thermal phase continuously changes the spacing between successive entropy maxima. Secondly, the time-dependent denominator introduces an additional modulation of the oscillation amplitude. Consequently, as the black hole evolves, the temporal entanglement structure gradually deforms. This reflects the changing thermodynamic state of the horizon. The plot for timelike entanglement entropy in this evaporating scenario is shown in Figure \ref{dynamic tEE}.
The condition for the extrema of the dynamical timelike entanglement entropy follows directly from the oscillatory phase. The maxima satisfy
\begin{equation}
\int_{0}^{t_n} \kappa(t')\,dt' = \Phi(t_n) =(2n+1)\pi, \qquad n=0,1,2,\cdots \label{eq:dynamicPage}
\end{equation}
The above equation (\ref{eq:dynamicPage}) provides a natural generalization of the Page-like times introduced in the static geometry. In the limit of constant surface gravity, $\kappa(t)=\kappa$, the accumulated phase reduces to $\Phi(t)=\kappa t$, and Eq.(\ref{eq:dynamicPage}) reproduces the static result,
\begin{equation}
t_n= \frac{(2n+1)\pi}{\kappa}.
\end{equation}
Unlike the static case, we get the Page like times as non-uniformly spaced. This demonstrates that the temporal entanglement structure continuously adapts to the evolving geometry of the Schwarzschild--AdS black hole.  The numerical evolution allows us to examine several important features that are absent in the static Schwarzschild-AdS geometry. In particular, we examine how the oscillation frequency, the amplitude of the timelike entanglement entropy, and the locations of successive Page-like times evolve as the black hole loses mass through Hawking radiation. By comparing the dynamical entropy with its static counterpart, we quantify the deformation of the temporal correlation structure induced by black hole evaporation and identify the characteristic signatures associated with the evolving Schwarzschild-AdS spacetime.

An important feature of the dynamical timelike entanglement entropy is that it retains a memory of the entire evaporation history through the accumulated thermal phase. Unlike the static case, where the entropy depends only on the constant surface gravity, the dynamical entropy depends on the integral of the time-dependent surface gravity over the complete evolution of the black hole. Consequently, the temporal correlations at a given instant are determined not only by the instantaneous thermodynamic state of the black hole but also by its previous evolution. This history-dependent behavior provides a richer characterization of temporal quantum correlations and suggests that the dynamical timelike entanglement entropy may serve as a complementary probe of quantum information flow during black hole evaporation. 
\begin{figure}
    \centering
    \includegraphics[width=0.5\linewidth]{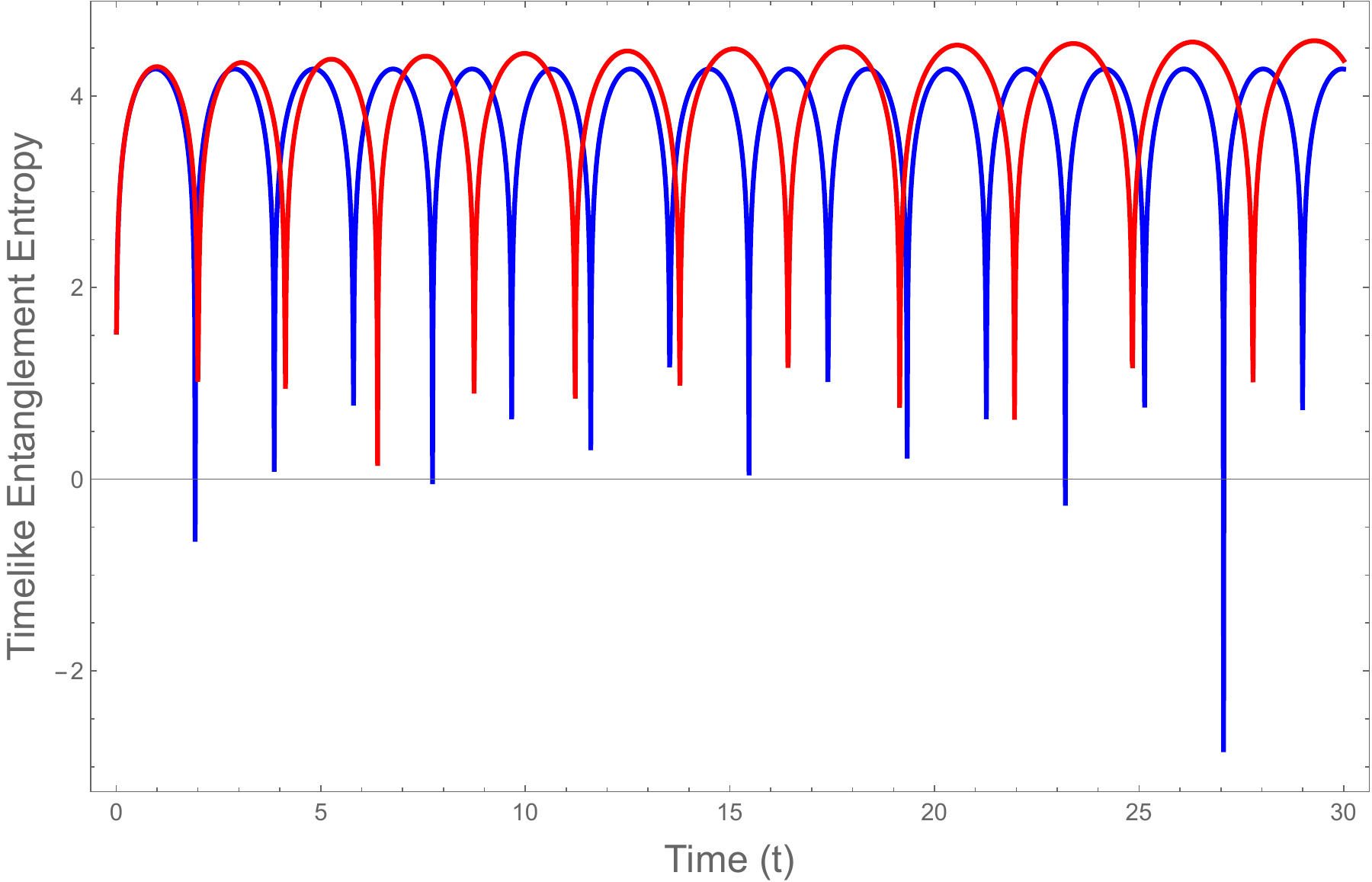}
 \caption{Comparison between the timelike entanglement entropy of the static (blue) and evaporating (red) Schwarzschild--AdS black holes.}
    \label{comparison}
\end{figure}
Figure~\ref{comparison} compares the timelike entanglement entropy obtained for the static Schwarzschild--AdS geometry with that corresponding to the evaporating background. The static curve (blue solid line) is obtained by assuming a constant surface gravity. On the other hand, the dynamical curve (red solid line) incorporates the time-dependent accumulated thermal phase together with the evolving surface gravity resulting from Hawking evaporation.
The numerical results demonstrate that the oscillatory behaviour of the timelike entanglement entropy persists throughout the evaporation process. However, the evolution is no longer governed by a constant oscillation frequency. Instead, the accumulated phase given by equation (\ref{eq:accumulatedphase}) continuously modifies the temporal structure of the correlations. This leads to  a gradual deformation of the oscillatory pattern.
\par
Compared with the static Schwarzschild--AdS background, two distinct effects become apparent. First, the oscillation peaks of the dynamical entropy exhibit a progressive temporal delay. This phase lag originates from the decrease of the instantaneous surface gravity during most of the evaporation process. Since the accumulated phase grows more slowly than the equilibrium phase $\kappa t$, the extrema of the timelike entanglement entropy occur at increasingly later times. This reflects the influence of the evolving black-hole geometry on the temporal correlation structure.
Secondly, the maxima of the dynamical entropy become slightly larger than those of the static configuration. This enhancement follows directly from the prefactor $1/\kappa(t)^2$ appearing inside the logarithm. As the surface gravity decreases during evaporation, the corresponding increase of the prefactor leads to larger peak values of the timelike entanglement entropy. Consequently, Hawking evaporation influences not only the temporal location of the entropy extrema but also the strength of the temporal quantum correlations.
Overall, the numerical analysis confirms that Hawking evaporation does not destroy the oscillatory nature of the timelike entanglement entropy. Instead, the evolving surface gravity continuously reshapes the temporal correlation structure through the accumulated thermal phase, leading to both a phase delay and an enhancement of the oscillation maxima. These features constitute the characteristic signatures of dynamical timelike entanglement entropy in evaporating Schwarzschild--AdS spacetime.

\section{Discussion and Conclusion}

In this work, we have developed a dynamical formulation of tEE for an evaporating Schwarzschild--AdS black hole, grounded in a first-principles derivation of the accumulated thermal phase. This accumulated thermal phase governs the time dependent evolving horizon. We started with the static Schwarzschild--AdS geometry, where the timelike entanglement entropy exhibits a periodic oscillatory structure which is governed by the constant surface gravity, with a corresponding sequence of uniformly spaced Page-like times. Coupling the black hole to an external absorptive bath \cite{almheiri2020page} allowed the black hole to evaporate through an effective Stefan--Boltzmann description of Hawking radiation, rendering the horizon radius and surface gravity explicitly time dependent.

To generalize the static construction to this evolving background, we derived the accumulated thermal phase $\Phi(t)=\int_0^t\kappa(t')\,dt'$ from the near-horizon Rindler structure of the evolving horizon directly. This construction rests on a single adiabatic ansatz, controlled by the parameter $\mathcal A(t)=|\dot\kappa(t)|/\kappa(t)^2$. We verified numerically that it remains small throughout the regime considered.
\par
The central result of this work is that the resulting dynamical timelike entanglement entropy is intrinsically history dependent: unlike the static case, where the entropy depends only on the instantaneous, constant surface gravity, the dynamical entropy at any instant $t$ depends on the entire integrated thermal history of the black hole up to that time, through $\Phi(t)$. Consequently, two evaporating black holes sharing the same instantaneous surface gravity but differing in their prior evaporation history would, in general, exhibit different timelike
entanglement entropy at that instant. This history dependence is the qualitatively new feature introduced by evaporation and has no counterpart in any static treatment of timelike entanglement entropy. It manifests concretely in two ways: the Page-like times, rather than remaining uniformly spaced as in the static case, become non-uniformly distributed in physical time, tracking the full accumulated phase rather than the instantaneous surface gravity; and the oscillation maxima of the entropy are modulated in amplitude as the surface gravity evolves, reflecting the changing thermodynamic state of the horizon at each instant while still encoding its earlier evolution.

We regard this history-dependent structure as the primary physical content of the dynamical timelike entanglement entropy constructed here: it suggests that tEE, unlike the standard (spacelike) entanglement entropy of Hawking radiation, may serve as a probe not only of the instantaneous state of an evaporating black hole but of its entire thermal history, offering a genuinely new observable window into black hole evaporation beyond the standard Page curve and island constructions.

The present analysis has been carried out within the semiclassical, quasi-static approximation, using an effective Stefan--Boltzmann model of Hawking evaporation where we neglected greybody factors, higher-order backreaction and quantum-gravitational corrections not incorporated.
Although these effects may quantitatively modify the results presented here, the general framework - and in particular the history-dependent character of the dynamical entropy is expected to persist beyond this approximation, and provides a useful starting point for future investigations.

The formalism presented here naturally admits several extensions. The most immediate direction is the study of charged and rotating AdS black holes, including Reissner--Nordstr\"om--AdS, Kerr--AdS and Kerr--Newman--AdS spacetimes, where additional thermodynamic parameters are expected to
produce richer temporal entanglement structures. It would also be interesting to investigate the influence of black-hole phase transitions on the dynamical timelike entanglement entropy, and to explore possible connections between temporal entanglement, holographic complexity, and quantum extremal surfaces. Another promising direction is to incorporate more realistic evaporation models, including greybody factors and backreaction beyond the adiabatic approximation. Such investigations may provide deeper insight into the role of temporal quantum correlations in black-hole thermodynamics and quantum information.
%

\end{document}